\documentclass[aps,prl,preprint,groupedaddress,showpacs]{revtex4-1}

\usepackage{color}
\usepackage{graphicx}
\usepackage{times}
\usepackage{amssymb}
\usepackage{multirow}
\usepackage{amsmath}
\usepackage{mathtools}
\usepackage{siunitx}
\usepackage{epstopdf}

\usepackage[normalem]{ulem}

\newcommand{\BSCCO}{Bi$_{2}$Sr$_{2}$CaCu$_{2}$O$_{8+x}$}
\newcommand{\ReSigma}{$\mathrm{Re}\ \Sigma$}
\newcommand{\ImSigma}{$\mathrm{Im}\ \Sigma$}
\newcommand{\EF}{$E_\mathrm{F}$}
\newcommand{\EBin}{$E_{\mathrm{bin}}$}
\newcommand{\taushort}{\tau_\mathrm{short}}
\newcommand{\taulong}{\tau_\mathrm{long}}
\newcommand{\Ebin}{\EBin}
\newcommand{\kk}{\mathbf{k}}

\begin{document}


\title{Energy dissipation in the time domain governed by bosons in a correlated material}
\pacs{abcd}


\author{J.~D.~Rameau,$^{1}$ S.~Freutel,$^{2}$  A.~F.~Kemper,$^{3,4}$ M.~A.~Sentef,$^{5,6}$ J.~K.~Freericks,$^{7}$ I.~Avigo,$^{2}$ M.~Ligges,$^{2}$ L.~Rettig,$^{2,\dagger}$ Y.~Yoshida,$^{8}$ H.~Eisaki,$^{8}$ J.~Schneeloch,$^{1}$ R.~D.~Zhong,$^{1}$ Z.~J.~Xu,$^{1}$ G.~D.~Gu,$^{1}$ P.~D.~Johnson,$^{1}$ U.~Bovensiepen$^{2}$}

\affiliation{$^{1}$Brookhaven National Laboratory, Upton, New
York, 11973, USA} \email[Direct correspondence
to:]{jrameau@bnl.gov} \affiliation{$^{2}$Faculty of Physics and
Center for Nanointegration Duisburg-Essen (Cenide), University
Duisburg-Essen, Lotharstrasse 1, 47057 Duisburg, Germany}
\affiliation{$^{3}$Department of Physics, North Carolina State
University, Raleigh, NC  27695, USA} \affiliation{$^{4}$Lawrence
Berkeley National Laboratory, 1 Cyclotron Road, Berkeley, CA
94720, USA} \affiliation{$^{5}$HISKP, University of Bonn, 53115
Bonn, Germany} \affiliation{$^{6}$Max Planck Institute for the
Structure and Dynamics of Matter, Center for Free Electron Laser
Science, 22761 Hamburg, Germany} \affiliation{$^{7}$Department of
Physics, Georgetown University, Washington, DC 20057, USA}
\affiliation{$^{8}$National Institute of Advanced Industrial
Science and Technology, Tsukuba, Ibaraki 305-8568, Japan}
\affiliation{$^{\dagger}$Current address: Swiss Light Source, Paul
Scherrer Institute, 5232 Villigen-PSI, Switzerland}



\date{\today}

\begin{abstract}
In complex materials various interactions play important
roles in determining the material properties. Angle Resolved
Photoelectron Spectroscopy (ARPES) has been used to study these
processes by resolving the complex single particle self energy
$\Sigma(E)$ and quantifying how quantum interactions modify
bare electronic states. However, ambiguities in the measurement of
the real part of the self energy and an intrinsic inability to
disentangle various contributions to the imaginary part of the
self energy often leave the
implications of such measurements open to debate. Here we employ a
combined theoretical and experimental treatment of femtosecond
time-resolved ARPES (tr-ARPES) and show how measuring the
population dynamics using tr-ARPES can be used to separate
electron-boson interactions from electron-electron interactions.
We demonstrate the analysis of a well-defined electron-boson
interaction in the unoccupied spectrum of the cuprate \BSCCO
characterized by an excited population decay time
that maps directly to a discrete component of the equilibrium self
energy not readily isolated by static ARPES experiments.
\end{abstract}

\maketitle

\section*{Introduction}

A host of interactions play a role in
determining the properties of complex quantum materials. Foremost among these are
interactions among the electron quasiparticles, and
between electron quasiparticles and bosonic excitations such as
phonons and spin fluctuations. Quantifying these interactions is
essential to understanding the fundamental and emergent phenomena
in complex materials. In particular, in strongly correlated
materials the phenomenology of these interactions remains vital to
understanding the behavior of charge and spin density waves,
electrical conductivity anomalies and unconventional
superconductivity. However, the analysis is challenging because
different experimental methods can easily lead to
different conclusions. The difficulty primarily arises because the
interactions overlap in energy, making them difficult to
disentangle.

Interactions quite generally give rise to the quasiparticle
lifetime, or equivalently an excitation energy linewidth. These
lifetimes are described by the energy dependent imaginary part of
the quasiparticle self-energy, \ImSigma($E$), which
represents the effect of the interactions on the quasiparticle at energy $E$ and
determines various properties of the material. Each of the
interactions has a specific contribution to \ImSigma\, which in equilibrium
add to result in a total linewidth according to Matthiessen's rule.  This superposition of
single particle self energies complicates the disentanglement of the individual interactions, which is desirable
to be able to understand the electron-boson (e-b) and electron-electron (e-e)
interactions.

Quasiparticle
lifetimes or their mass renormalizations can
in some cases be directly measured using angle-resolved
photoemission spectroscopy (ARPES) by studying the linewidth
$\Gamma(E)$
or effective dispersion $E(k)$, obtaining the imaginary or
real parts of the self-energy (\ImSigma\ and \ReSigma),
respectively
\cite{lanzara_evidence_2001,lee_aspects_2007,johnston_evidence_2012,valla_many-body_1999, schaefer_electronic_2004}
%
Recently, femtosecond time-resolved ARPES (tr-ARPES)
\cite{petek_femtosecond_1997,perfetti_ultrafast_2007,graf_nodal_2011,cortes_momentum-resolved_2011,bovensiepen_elementary_2012,hellmann_time-domain_2012}
 has come to the forefront as another method for studying the quasiparticle lifetimes through analyzing population dynamics,
and thus
ostensibly the interactions. In these experiments, an ultrafast
laser pump excites quasiparticles from their ground state, and
their subsequent energy- and momentum-resolved relaxation to equilibrium  is probed by analysis of the photoelectron spectrum
generated by a second, much weaker, time-delayed probe pulse in the UV or XUV
spectral
range.
The lifetime analysis of the laser excited population relaxation
by kinetic rate equations or density matrix formalism
\cite{petek_femtosecond_1997,ueba_theory_2007} has proven successful for quasiparticle
energies between approximately 0.5 to several eV in a limit where
relaxation is dominant at all times and quasiparticle lifetimes
can indeed be determined in the time domain.
\cite{petek_femtosecond_1997,knorren_dynamics_2000,echenique_decay_2004,kirchmann_quasiparticle_2010}

At energies close to the Fermi level $E_{\mathrm{F}}$ the
situation is more challenging because two effects become important
and hamper the lifetime determination.
There the various interactions do not simply add to determine the population
dynamics
\cite{yang_inequivalence_2015}. Here we propose to exploit this seeming complication
to separate the microscopic interactions of interest using
population dynamics observed in tr-ARPES.
Immediately after the pump, the primary electronic
excitations relax by e-e scattering, which
leads to a population redistribution of secondary electronic
excitations. Subsequently, coupling to phonons becomes in general the
dominant contribution
\cite{valla_many-body_1999,perfetti_ultrafast_2007}. As a result, the electron-phonon (e-p) coupling leaves its
fingerprint on the population dynamics.

In this work, we show that microscopic conclusions can be drawn from
this particular fingerprint. Using tr-ARPES, we observe
population relaxation dynamics which we describe by
interaction with a specific phonon after excitation by a light field.
We apply this analysis to the model conducting system \BSCCO (Bi2212),
which is a good candidate because it exhibits pronounced e-p scattering at discrete and well-known energies\cite{lanzara_evidence_2001,lee_aspects_2007,johnston_evidence_2012}.
We find quantitative
agreement between experiment and theory. By explicit inclusion of the pump
light field in our theoretical description as the primary electronic excitation mechanism, we avoid the
assumption of an initially thermalized electronic distribution which was
required in previous work \cite{perfetti_ultrafast_2007}.
The population dynamics are calculated using a numerical time-dependent Green's function approach.\cite{kemper_effect_2014}
The photoexcited electrons scatter on an ultrashort
time scale by e-e interactions. Because these
interactions conserve the total electronic energy in the electron
system, they cannot be responsible for the dissipation of the excess electronic energy. The dissipation is achieved
by electron-phonon (e-p) interactions which transfer the energy
from the electron system via the $\hbar\Omega\approx 75$~meV phonon to a heat bath
representing acoustic phonons. We validate this assertion by a
comparison between the tr-ARPES measurements and a theoretical
calculation which only includes e-p interactions in the
relaxation dynamics.

A measurement of population dynamics on
the copper-oxide material Bi2212 in its normal state is
used both because this material is known to host very strong
e-b interactions throughout its phase diagram~\cite{damascelli_rmp_2003,lanzara_evidence_2001,johnson_doping_2001}
and to demonstrate the ability of our approach to discern such
interactions even when obscured by other physics, as is the case
in the normal state of Bi2212.\cite{johnson_doping_2001}
The experimentally observed relaxation dynamics
exhibit two characteristic timescales, $\taushort(E)$ and
$\taulong(E)$, of which $\taushort(E)$ exhibits a sudden speedup at
energies above $+\hbar\Omega$ for the lowest pump fluence used.
The newly discovered short timescale $\tau_{short}$ is found to
account for a large part of the lifetime obtained from the
equilibrium line width $\tau_{QP}(E)$ measured near \EF\
along the zone diagonal. By
comparison to theory, we show that these time scales can arise
from the coupling to a single phonon, namely the 75 $\si{meV}$ $O$
breathing mode\cite{lanzara_evidence_2001,zhou_handbook_2007}. This discovery suggests a new use for
time-resolved measurements \textemdash as a selective probe of
e-b interactions in materials where their presence and strength are obscured
by other processes.

\begin{figure}
    a.
    \includegraphics[width=0.45\textwidth,clip=true,trim=0 0 0 0]{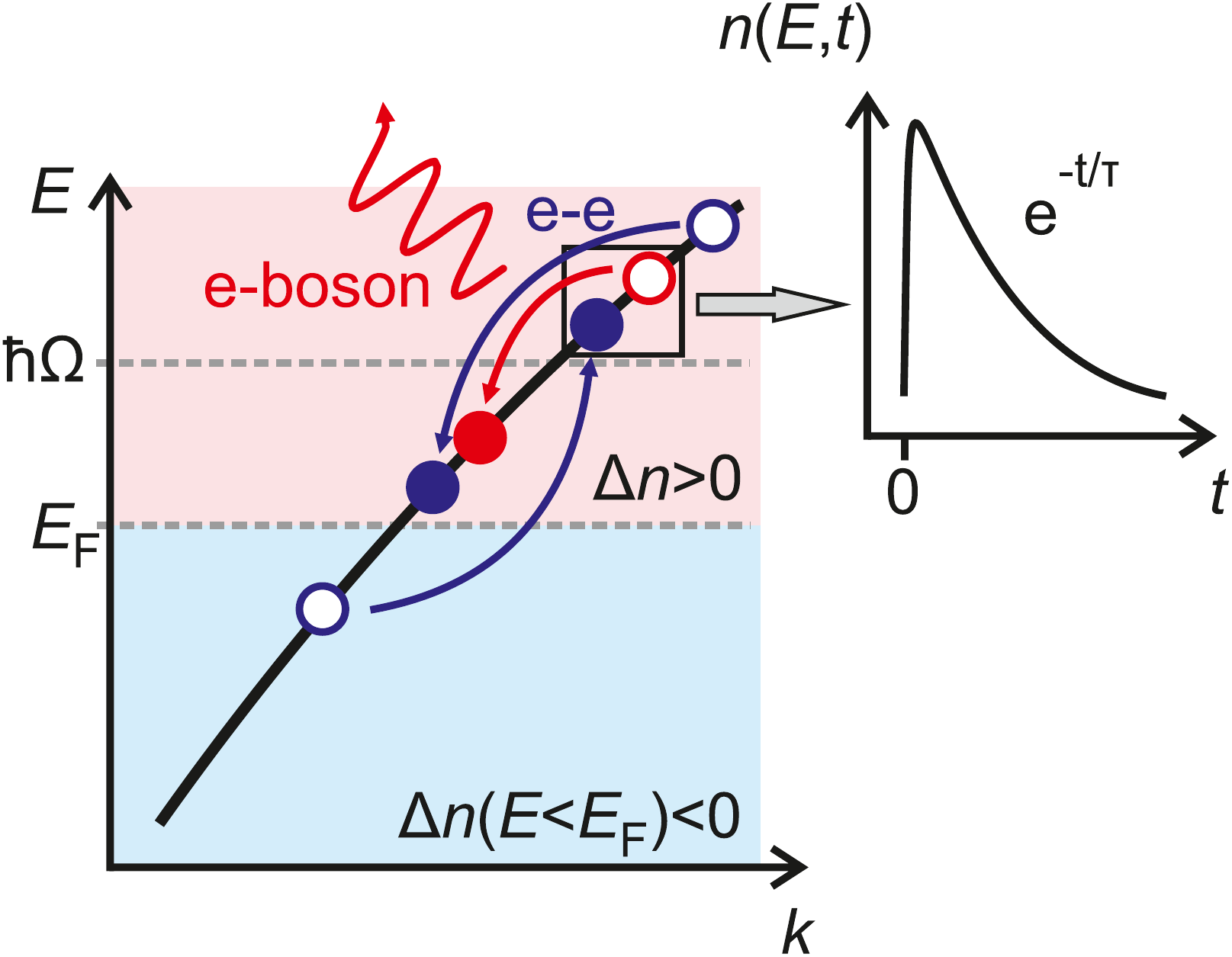}
    b.
    \includegraphics[width=0.45\textwidth]{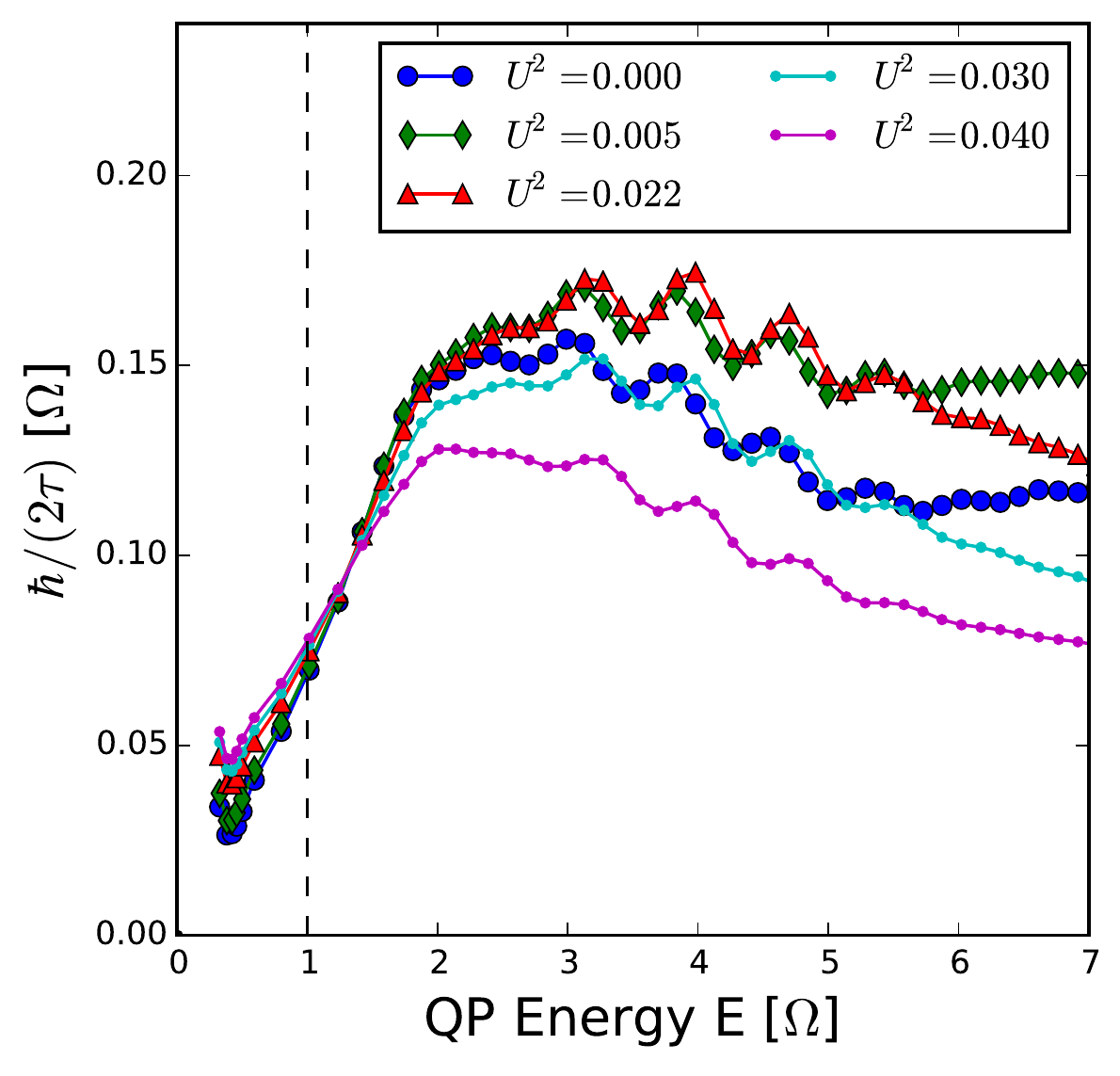}
    \caption{{\bf Time domain separation of electron-electron and electron-boson scattering.} a. Illustration of the dissimilarity between e-b and e-e scattering for population dynamics.
    Electron-electron scattering (blue arrows)
        by nature maintains the amount of energy in the electrons, limiting it to rapid redistribution of energy
	within the system. On the other hand, e-b (red arrows) scattering can carry energy out, leaving a measurable impression
	on the population dynamics. This causes the population dynamics to decay
	with the characteristics of the e-b scattering.
    b.
    Population decay rate extracted from the time-dependent density $n(E,t)$ along the zone diagonal $k_x=k_y$
    as a function of the quasiparticle energy $E$ for various Coulomb interaction
    strengths $U$. Here, $g^2=0.02$ (all interaction strengths are in eV$^2$) and the phonon frequency $\Omega=0.1$ eV.
    }.
    \label{fig:scatt_compare}
\end{figure}

We begin with some general remarks regarding the dynamics of
population
distributions when driven out of equilibrium. The pump deposits
energy into the electrons. If e-e interactions are the only interactions present, the excess
energy must remain in the electronic system, and although the
scattering can cause a rapid change in the energy and momentum
dependent population, it can only thermalize the electrons at a
higher temperature determined by the absorbed energy. However, a coupling to a
bosonic bath of phonons can draw out the energy by transferring it
to the lattice, which has a larger heat capacity than the electronic system.  The
determining factor for the dynamics is then the rate at which
e-p scattering dissipates the energy, which
depends critically on the e-p coupling strength and the phonon
frequencies.

The full dynamics of this process, including e-e and e-p interactions, involves a complex interplay
beyond this simple model. Energy
relaxation of the photoexcited carrier population is expected to
occur on three distinct timescales~\cite{kemper_effect_2014}
(i) electron population redistribution on a femtosecond timescale mediated by
e-e scattering;
(ii) transfer of energy from the electrons to strongly coupled phonons
on a tens- to hundreds of femtoseconds timescale; and
(iii) full
system thermalization on picosecond or longer timescales, on which the excited population of the strongly coupled phonons decays through anharmonic
coupling or transport.
In Fig.~\ref{fig:scatt_compare} we illustrate processes (i) and (ii), which are the
relevant ones for the short term dynamics.  The measured and calculated time traces typically
show a rise at $t=0$, followed by a decay. The dynamics are determined
by a combination of processes (i) and (ii), where e-p and e-e scattering cooperate in a non-additive fashion
to equilibrate the electrons with each other as well as with the phonon bath.
However, only e-p interactions can carry energy away from the electrons in a non-magnetic system.

To show that the
e-p scattering is the determining factor in the return
to equilibrium, we performed simulations of the pump-probe process
using non-equilibrium Keldysh Green's functions. We treat the
e-p and e-e interactions self-consistently, with both
interactions being evaluated at second order in perturbation
theory. Here, we go beyond previous
work\cite{kemper_mapping_2013,sentef_examining_2013,kemper_effect_2014} and include the
effects of e-e scattering in second order
self-consistently. We calculate the time-dependent density
$n(\kk,t)$ along the zone diagonal after excitation by a short 1.5~eV pump pulse for a
simple nearest-neighbor tight-binding model and then fit the time
traces with a single exponential. To illustrate that the resulting
decay rates reflect the e-p scattering even when e-e scattering
is present, we obtain the decay rates for various e-e scattering
strengths ($U$). Fig.~\ref{fig:scatt_compare}b shows the decay
rates obtained for quasiparticles at energy $E$ above $E_{\mathrm{F}}$. The rates
without e-e scattering show a step at the phonon
frequency\cite{sentef_examining_2013,kemper_effect_2014}. Strikingly, even when the e-e
scattering is increased beyond the e-p scattering strength ($U^2
> 0.02$), the step remains visible in the data. We also
note that the high-energy decay rate is smaller when e-e
scattering is present, suggesting that there is some competition
between the two processes.
This originates from the fact that the population dynamics is a fundamentally different
quantity than the dynamics (and therefore lifetime) of a singly-excited quasiparticle.
The latter is determined by the wavefunction decay, and the former by the interactions the quasiparticles
with each other and with external phonons.

\begin{figure}[!]
  \includegraphics[width = 8.6 cm]{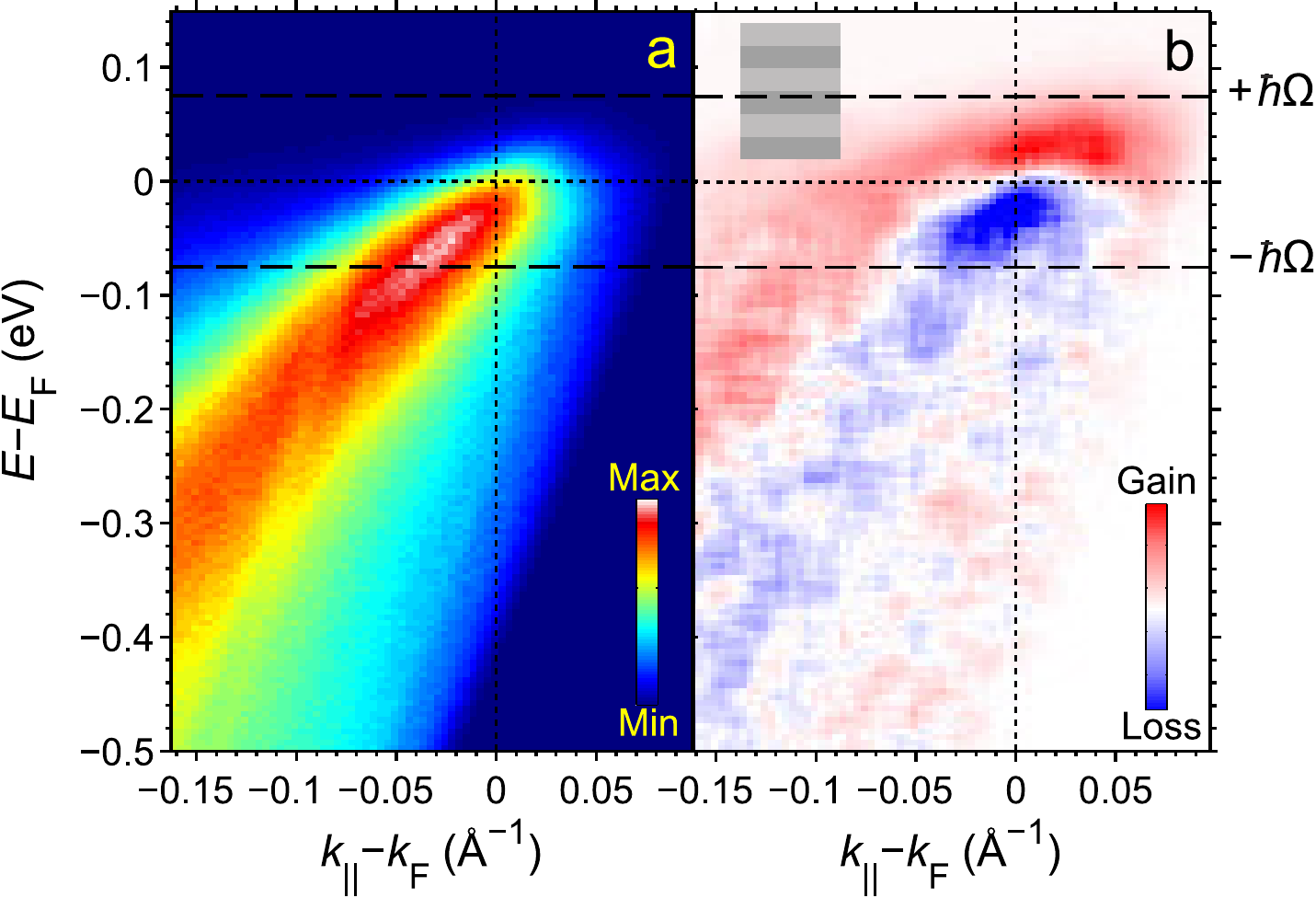}\\
  \caption{{\bf Equilibrium and excited photoelectron spectra.} (a) Photoelectron intensity $I(E,k_{||},t_{\mathrm{eq}})$ along the node of Bi2212 at equilibrium, $t_{\mathrm{eq}}$ = -5 ps and equilibrium temperature $T$ = 120 K. b) Difference spectrum $\Delta I=I(E,k_{\parallel},t)-I(E,k_{\parallel},t_{\mathrm{eq}})$ for $\Phi = 35 \mathrm{\mu}$J$\mathrm{\cdot}$cm$\mathrm{^{-2}}$, $t$ = 50 fs. Light and dark gray bars in panel b) depict energy bins $\Delta E=20$ $\si{meV}$ centered at $E_{\mathrm{bin}}$ used to evaluate $I_{\mathrm{norm}}(E_{\mathrm{bin}},t)$.}\label{fig1}
\end{figure}

\section*{Experiment}

We now demonstrate that this analysis can be applied in real
materials by making a quantitative comparison between theoretical
simulations of the type shown above and a tr-ARPES experiment. A
typical equilibrium spectrum $I(E,k_{\parallel},t_{eq})$ along the
nodal direction of Bi2212 is shown in Fig. \ref{fig1}a. Here $I$ is
photoelectron intensity and $t_{eq}=-5$ ps precedes the pump
pulse. Though the dispersion kink is difficult to observe without
spectral deconvolution at this resolution~\cite{rameau_photoinduced_2014},
the increase in coherence (or decrease in linewidth) of the states
between $-\hbar\Omega$ and \EF\ is still clear. In the
e-b picture, this is due to a decrease in scattering
rate for energies below the boson energy $\hbar\Omega$.
Fig.~\ref{fig1}b shows the difference spectrum $\Delta
I(E,k_{\parallel},t)=I(E,k_{\parallel},t)-I(E,k_{\parallel},t_{eq})$
where $t=50$ fs, and at incident pump fluence $\Phi = 35 \mathrm{ \mu
J\cdot cm^{-2}}$.
For this fluence, by $t=50$ fs electrons excited above the equilibrium kink energy $+\hbar\Omega$ and holes injected below
$-\hbar\Omega$ have largely relaxed, while those within the boson window $-\hbar\Omega<E<+\hbar\Omega$ persist.
Considering that the pump photon energy of 1.5 eV (see Methods section for details) is well in excess of this $2\hbar\Omega$ range,
our tracking of the excitations in the vicinity of \EF\ implies we are primarily observing secondary electrons and holes and their
relaxation towards \EF, rather than the initial excitation~\cite{yang_inequivalence_2015}.
Indeed, Fig. \ref{fig1}b shows the largest pump-induced intensity changes within the boson window.
The existence of this pileup in excited carriers at finite $t$ means the relaxation times for carriers outside the boson window must be
considerably shorter than inside since the former carriers have already relaxed by this time.
\cite{graf_nodal_2011,sentef_examining_2013}
Below, we show that the significant change in relaxation times in the unoccupied part of the spectrum around $+\hbar\Omega$
results from the sudden change in phase space with increasing $E$ for the inelastic decay of excited carriers by emission of a
single predominant bosonic mode.
This conclusion is based upon quantitative agreement between our experiment and the theory of fluence-dependent,
boson-derived relaxation times which do not account for e-e scattering.

\begin{figure}[!]
  \includegraphics[width = 8.6 cm]{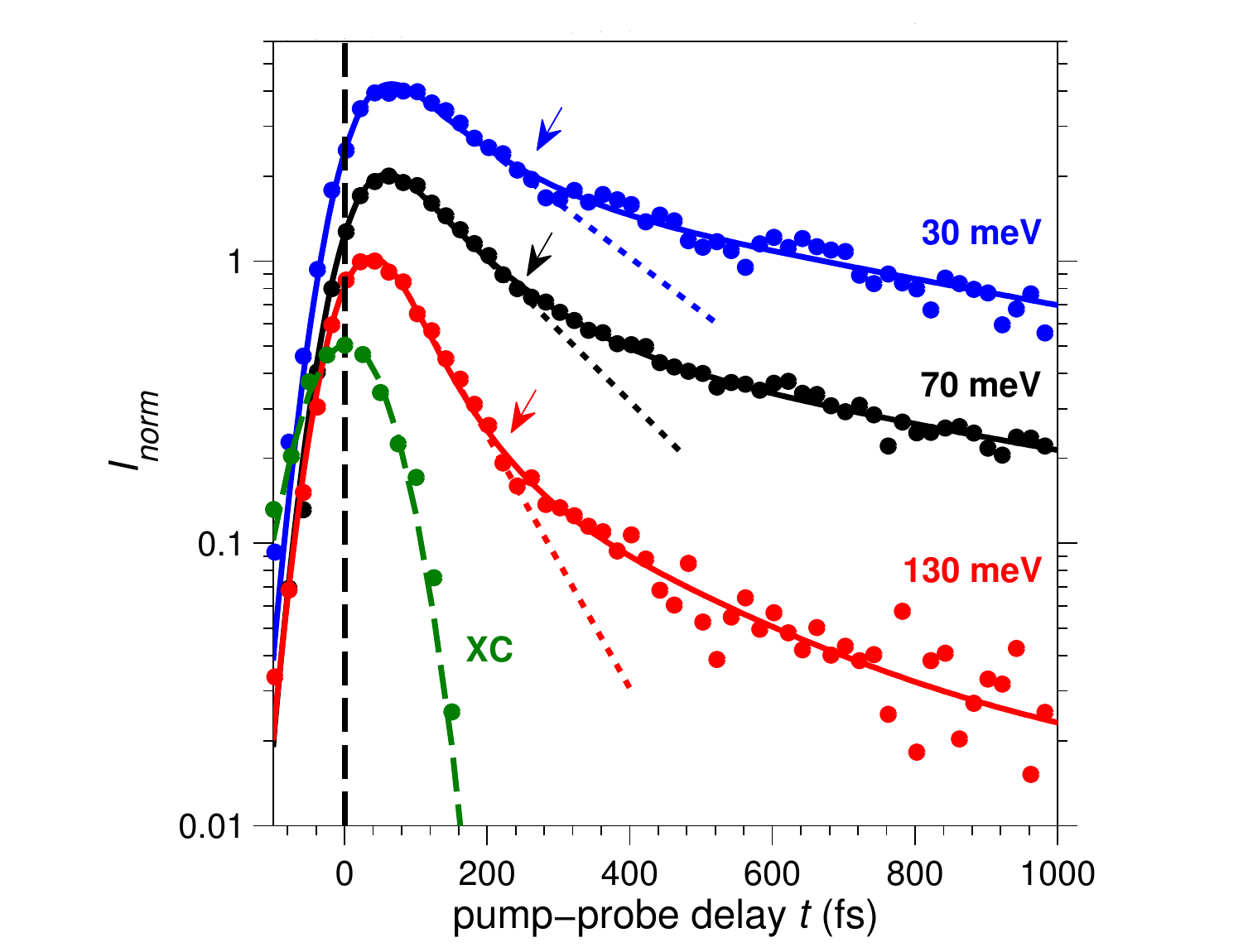}\\
  \caption{{\bf Time-resolved population decay.} Intensities $I_{\mathrm{norm}}(E_{\mathrm{bin}},t)$ at $\Phi = 315$ $\mathrm{\mu J\cdot cm^{-2}}$ for selected binding energies $E_{\mathrm{bin}}$ = 30 meV (blue circles), 70 meV (black circles) and 130 meV (red circles), respectively, on a logarithmic intensity scale. Solid lines are fits to Eq. \ref{model}. Data for each $E_{\mathrm{bin}}$ are offset by a factor of two for clarity. Dotted lines are guides to the eye extending through $I_{\mathrm{norm}}(E_{\mathrm{bin}},t)$ at times for which population decay is dominated by the scattering mechanism responsible for $\tau_{\mathrm{short}}$. Green points are from $I_{\mathrm{norm}}(E,t)$ integrated over all $k_{\parallel}$ and $1.45 < E < 1.5$ eV for $t<200$ fs, representing mainly the cross correlation (XC) time of pump and probe pulses. The Gaussian fit to the green points (green dashed line) yields the XC width. Arrows denote time at which $\tau_{long}$ dominates $I_{\mathrm{norm}}(t)$.}\label{fig2}
\end{figure}

To quantify these effects, we introduce a procedure to energy-resolve the photoexcited population relaxation
times $\tau$ above \EF\ in a manner that allows comparison to the equilibrium quasiparticle lifetime $\tau_{\mathrm{QP}}$
determined by analysis of energy- and momentum distribution curves (EDCs and MDCs, respectively) in equilibrium ARPES.
This is accomplished by dividing the tr-ARPES spectrum into discrete energy bins of equal width $\Delta E = 20$ meV
(Fig.~\ref{fig1}b).
Integrating the intensity within bins over $k_{\parallel}$ and $\Delta E$ results in time- and energy-dependent intensities
$I(E_{\mathrm{bin}},t)$, where $E_{\mathrm{bin}}$ is taken at the bin centers.

Our subsequent analysis follows from the observation of a
pronounced biexponential decay of photoexcited carrier populations
with respect to \EBin\ and $\Phi$. Typical energy-resolved data for
$\Phi =$ 315 $\mathrm{\mu J\cdot cm^{-2}}$ is shown in Fig.
\ref{fig2}. The normalized intensities $I_{\mathrm{norm}}\equiv
I(E_{\mathrm{bin}},t)/\mathrm{max}[I(E_{\mathrm{bin}},t)]$ are
seen to decay with clearly separable short and long time
components ($\taushort$(\EBin) and $\taulong$(\EBin)), the former
lasting several hundred femtoseconds.
It is
this direct observation of two exponential time scales within the
population decay, rather than one, that allows the present
   analysis of the
electron
population decay mechanisms in the cuprates~\cite{yang_inequivalence_2015}. Fits to
$I_{\mathrm{norm}}(E_{\mathrm{bin}},t)$ at fixed \EBin in Fig.
\ref{fig2}, as well as all subsequent fits used to determine
$\tau_{\mathrm{short}}$ and $\tau_{\mathrm{long}}$, respectively,
are performed using the function
\begin{equation}\label{model}
    I_{\mathrm{norm}}(t) = \Theta(t)(A_{\mathrm{short}}e^{-\frac{t}{\tau_{\mathrm{short}}}}+A_{\mathrm{long}}e^{-\frac{t}{\tau_{\mathrm{long}}}})\otimes R_{\mathrm{t}}
\end{equation}
where $\Theta(t)$ is the Heaviside function, $A$'s are intensity amplitudes and $R_{\mathrm{t}}$ is a unit-normalized Gaussian of width equal to the cross-correlation (XC) of pump and probe responses at the sample surface.

\begin{figure}
  \includegraphics[width = 12.9 cm]{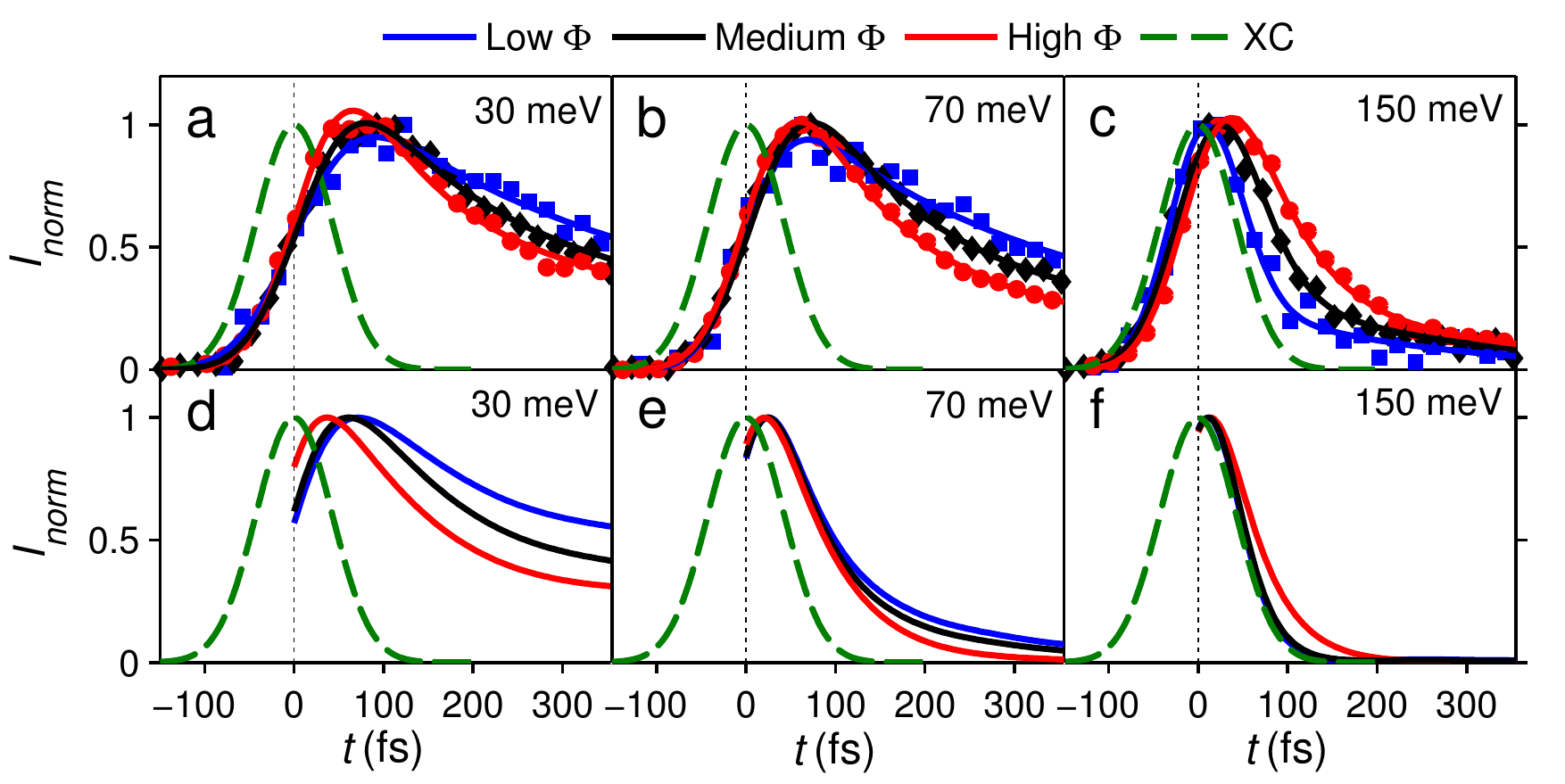}\\
   \caption{{\bf Experimental and theoretical fluence-dependent relaxation.} a-c) $I_{\mathrm{norm}}$ for $E_{\mathrm{bin}}=$ 30, 70 and 130 meV, respectively for $\Phi = 35$ (blue squares), $105$ (black diamonds) and $315$ (red circles) $\mathrm{\mu J\cdot cm^{-2}}$. Solid lines in these panels are fits to Eq.~\ref{model}. d-f) theoretical $I_{\mathrm{norm}}$ for the same $E_{\mathrm{bin}}$ as a-c for field strengths of 0.05 (blue), 0.15 (black) and 0.5 (red) in units of V/$a$ where $a$ is the Cu-Cu lattice constant. a-f) Green dashed lines show the Gaussian XC.}\label{fig3}
\end{figure}

To facilitate examination of both the energy and fluence dependencies of $\taushort$ and its comparison to theory we show
representative $I_{\mathrm{norm}}(t)$ for several \EBin\ and $\Phi$ together with the fits in Fig.~\ref{fig3}a-c.
The main trends in the data are the overall decrease in $\taushort$ with increasing $E_{\mathrm{bin}}$ for all fluences
and the changes in the ordering of the curves for $t>100$ fs, or equivalently the magnitude of $\taushort$.

\begin{figure}
  \includegraphics[width = 8.6 cm]{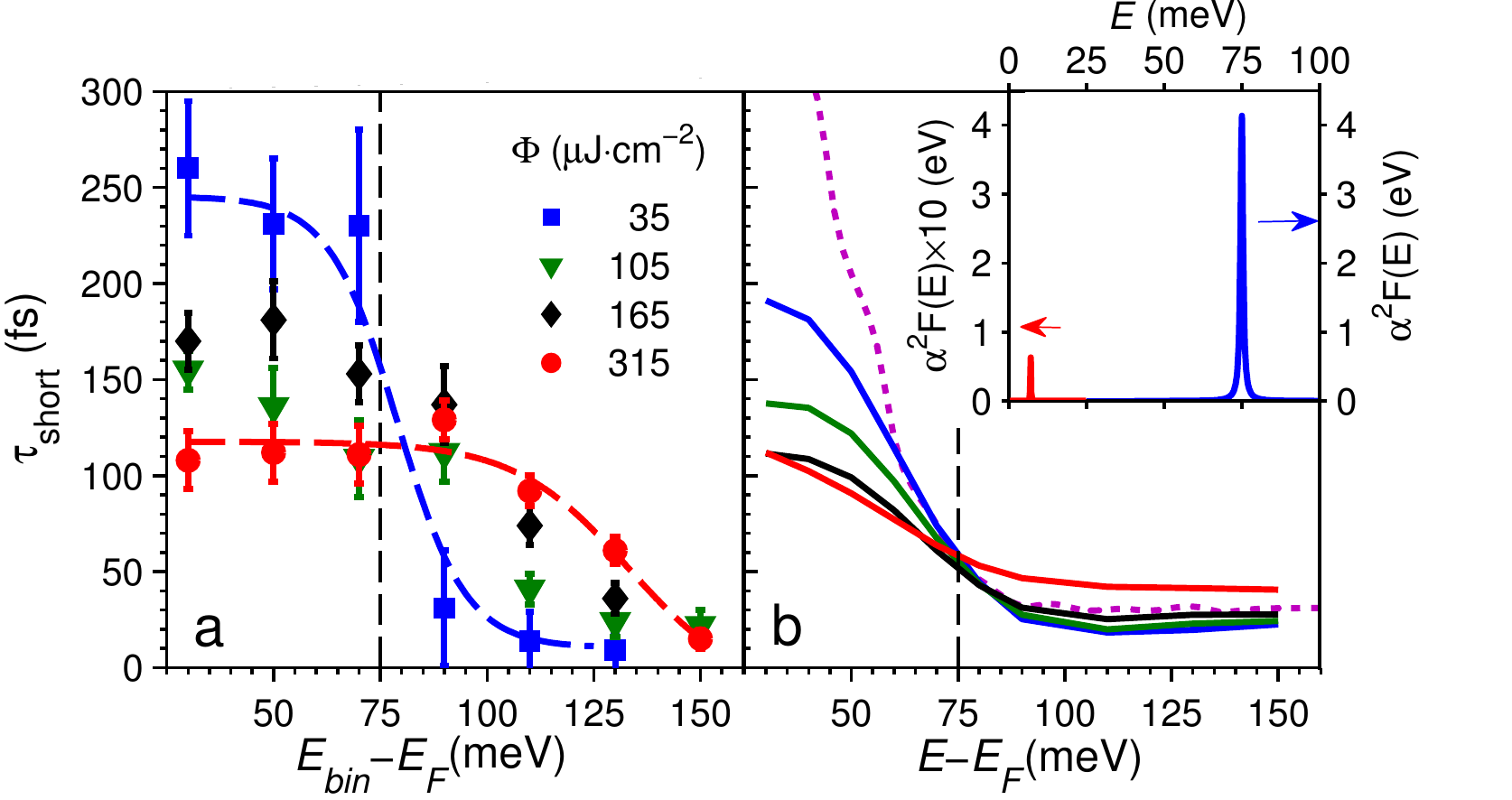}\\
  \caption{{\bf Fingerprint of electron-boson scattering.} a) Experimental $\tau_{\mathrm{short}}(E_{\mathrm{bin}})$ for $\Phi=$ 35 (blue squares), 105 (green triangles), 165 (black diamonds) and 315 (red circles) $\mathrm{\mu}$J$\mathrm{\cdot}$cm$\mathrm{^{-2}}$, respectively. Dotted blue and red lines are guides to the eye for the 35 and 315 $\mathrm{\mu}$J$\mathrm{\cdot}$cm$\mathrm{^{-2}}$ data, respectively. b) Theoretical $\tau_{\mathrm{short}}(E_{\mathrm{bin}})$ for field strengths of 0.05 (blue), 0.15 (green), 0.25 (black) and 0.50 (red) and equilibrium $\tau_{\mathrm{QP}}=\hbar/2\mathrm{Im}\ \Sigma$ (purple). The mode energy at 75 meV is marked by the dashed black lines. Inset) (blue, right axis) Theoretical $\alpha^{2}F(E)$ with e-b coupling constant $\alpha$ and dispersion $F(E)$. (red, left axis) The low energy lifetime dampening mode at 15 meV is magnified by 10x for clarity.
}\label{fig4}
\end{figure}

Plotting $\taushort$(\EBin) for each fluence (Fig. \ref{fig4}a), we observe a highly unusual relationship between the fluence
dependence of $\taushort$(\EBin) and the value of \EBin\ relative to $\hbar\Omega$.
At the lowest fluence $\taushort$(\EBin) shows a pronounced step at \EBin $\sim\hbar\Omega$ with much shorter times above
$+\hbar\Omega$ versus below.
As pump fluence is increased,  relaxation times below $\hbar\Omega$ decrease while those above $\hbar\Omega$ increase.
The step in $\taushort$ at $\hbar\Omega$ for 35 $\mathrm{\mu J\cdot cm^{-2}}$ is reminiscent of the step in $\hbar/\tau_{QP}$ (shown in Fig.~\ref{fig4}b)) accompanying boson-induced mass renormalizations observed in ARPES~\cite{valla_evidence_1999,valla_many-body_1999,lanzara_evidence_2001}.
In ARPES, a single photohole injected above a bosonic mode energy has a shorter lifetime than one injected below because an additional interaction channel is open to the higher energy state. In energy- and momentum distribution curves (EDC's and MDC's, respectively)
this phenomenon is observed as a step, centered at the mode energy, in the Lorentzian widths of the states. Likewise, an electron photoexcited above a bosonic mode's energy will decay faster than one excited to empty states below because the additional channel for energy loss due to boson emission is open to it.\cite{kemper_effect_2014}

The fluence dependence of these phenomena indicate a strong connection between the step in $\taushort$ observed above $E_{F}$ and the nodal kink observed below. It is the process of ``filling" states within the boson window~\cite{kemper_effect_2014} that leads to the characteristic speeding-up and slowing-down observed here as well as the weakening of the 70 $\si{meV}$ kink structure observed in previous experiments~\cite{rameau_photoinduced_2014,zhang_ultrafast_2014}. As $\Phi$ is increased the dynamics of photoexcited carriers change from a regime reflecting essentially single-particle physics to one further characterized by the dynamics of populations of interacting, hot particles. The parallel evolution, with increasing fluence, of the broadening step in $\taushort$ and the previously observed reduction in effective mass of the nodal kink below \EF\ therefore encourages a description of the former effects within the same theoretical framework as the latter. The absence of fluence dependence for $\taulong$(\EBin)
(see Methods)
points to its likely origin in the coupling of hot electrons to a broad spectrum of collective excitations such as acoustic phonon modes.

\subsection*{Comparison of Experiment to Theory}


With e-p scattering being the dominant mode of energy dissipation,
we now focus on a theoretical description of (ii), given experimentally by $\taushort$, and described by the Holstein model, where electrons are coupled to a single Einstein mode.
Since the main feature and overall scale is expected to be due to the phonon coupling as
discussed above, we neglect e-e scattering here.
Excited carriers relax via their coupling to this reservoir, which is assumed to have infinite heat capacity and remains unchanged.
The equilibrium $\tau_\mathrm{QP}(E)$ in Fig.~\ref{fig4}b is defined through the equilibrium line width due to a single well-defined phonon mode centered at $\hbar\Omega =$ 75 meV.
The dimensionless e-p coupling constant $\lambda$ is chosen to be $0.2$ to match the relaxation times of $\approx 20$ fs
at $E>\hbar\Omega$.
The same value for $\lambda$ was found to bound the coupling of hot electrons to a particular subset of hot phonons in earlier tr-ARPES experiments on Bi2212~\cite{perfetti_ultrafast_2007}, and is similar to that obtained from some equilibrium ARPES experiments\cite{lee_aspects_2007}.
The model also includes a weakly coupled mode at low energies to prevent an infinite $\tau_{\mathrm{QP}}$ within the phonon window ($-\hbar\Omega<E<+\hbar\Omega$).
It should be noted that the inclusion of e-e scattering
appears to decrease the size of the step in the scattering rates, slowing down the decay at energies above $\hbar\Omega$
as can be seen in Fig.~\ref{fig:scatt_compare}.
This could
lead to a lower apparent $\lambda$ in time-resolved experiments.

Representative $I_{\mathrm{norm}}(E_{\mathrm{bin}},t)$ extracted from theoretically generated time-dependent occupied spectral functions
are shown in Fig. \ref{fig3}d-f for field strengths comparable to the experiment. They successfully reproduce the reversal of the fastest $\tau_{\mathrm{short}}(E_{\mathrm{bin}})$ as a function of fluence for $E_{\mathrm{bin}}$ above as compared to below $+\hbar\Omega$.
Here, we focus on processes (i) and (ii), and ignore the long-time component, fitting the calculated
$I_{\mathrm{norm}}(E_{\mathrm{bin}},t)$ with a single exponential to produce
theoretical $\tau_{\mathrm{short}}(E_{\mathrm{bin}})$ for several fluences, Fig.~\ref{fig4}b.
The theoretical $\taushort$(\Ebin) exhibit the same step at $+\hbar\Omega$ as the experimental data in Fig.~\ref{fig4}a. The theory largely reproduces the increase in $\tau_{\mathrm{short}}(E_{\mathrm{bin}})$ for $E_{\mathrm{bin}}>+\hbar\Omega$ and decrease in $\tau_{\mathrm{short}}(E_{\mathrm{bin}})$ for
$0<E_{\mathrm{bin}}-E_{\mathrm{F}}<+\hbar\Omega$ as a function of increasing fluence.
%
The remarkable quantitative agreement between the measured and calculated $I_{\mathrm{norm}}(E_{\mathrm{bin}},t)$ within the phonon window, Fig.~\ref{fig3},  demonstrates that
scattering due to phonon emission is sufficient to describe the population dynamics measured with tr-ARPES,
allowing for the measurement of a single component of the equilibrium self-energy.

\section*{Discussion}

The experiment and theoretical calculations both capture the dynamics of energy dissipation through electron-phonon interactions.
The fingerprint of the e-p interaction through its particular phase-space restrictions is seen in the data through
both energy and fluence dependence of the population decay rates.

Understanding the detailed nature
of the equilibrium self-energy is thought to be vital to understanding a
host of phenomena in strongly correlated materials and beyond,
not least those derived from non-Fermi liquid properties that would be
expected to show up most strongly in electronic channels. In equilibrium ARPES,
it is difficult to disentangle phononic from other contributions to the self-energy,
especially when several different interactions are present.

The combination of electron-electron (in particular at higher temperatures), electron-boson, and electron-impurity scattering
\textemdash just to name a few \textemdash
collude and increase the overall linewidth, making it difficult to separate a single contribution.
With the techniques demonstrated here, we now have the ability to
isolate the phonon contributions directly from the total equilibrium self energy
through measurement of population dynamics with tr-ARPES with high temporal resolution
and subsequent comparison to ARPES data acquired with high energy resolution.

This finding has important ramifications beyond the present study.
Here, we have focused on a material where the presence of
a complex interplay of e-e and e-b interactions
is well known, but the general topic reaches much further.  This reach includes other strongly correlated materials
such as the Fe-based superconductors, heavy fermion compounds, and manganites, as well as more
weakly correlated materials in which strong e-p interactions can still occur.
Topological insulators are a field where the study of population dynamics is commonly used, and the methodology
presented here can aid in the interpretation and analysis of the results\cite{wang_measurement_2012,crepaldi_ultrafast_2012,hajlaoui_ultrafast_2012,sobota_ultrafast_2012,hajlaoui_tuning_2014,kim_robust_2014,cacho_momentum-resolved_2015}. Moreover, the particular relevance of bosonic mode couplings for excited state population dynamics which is of high interest, for instance, in light-harvesting applications. We point out that this relevance is often implicitly assumed, for example in recent ab initio computations of equilibrium self-energies, from which information about hot carrier dynamics for solar cells was inferred.~\cite{solarcalc} Our study provides a firm ground for such implicit assumptions, and the methods laid out by us can guide further efforts in this direction.

\section*{Methods}

\subsection*{Sample Preparation and Setup of tr-ARPES Experiment}
Our trAPRES experiment was performed on single crystals of optimally doped Bi2212 ($T_{c}$ = 91 K) grown by the floating zone method. $T_{c}$ was confirmed by SQUID magnetometry. Samples were cleaved \textit{in situ} at the lattice equilibrium temperature, 100 K, in a base vacuum of $5\times10^{-11}$ mbar. Pump pulses of 800 nm wavelength and 40 fs duration, at 250 kHz repetition rate, were produced by a Coherent RegA 9040 regenerative Ti:Sapphire amplifier. The 200 nm probe beam was produced as the fourth harmonic of part of the RegA's 800 nm fundamental using nonlinear crystals~\cite{perfetti_ultrafast_2007}. The pump-probe cross-correlation (XC, Gaussian full width at half maximum) was determined to be 100 fs at the sample surface. Photoelectron spectra with an overall energy resolution of 55 meV, set by the bandwidth of the laser pulses, were recorded using both angle integrating time-of-flight (TOF)~\cite{perfetti_ultrafast_2007} and position sensitive TOF (pTOF)~\cite{kirchmann_time--flight_2008} electron spectrometers. The $k$ resolution of the pTOF was 0.0025 $\mathrm{\AA^{-1}}$ and the angular acceptance of the TOF was $\pm3^{\circ}$.

\subsection*{Theoretical Modeling of the Data}

For the modeling, we use a tight binding parametrization of the one-band model for cuprates \cite{eschrig_effect_2003} with a filling of 0.42 per spin, i.e.~16$\%$ hole doping. The methodology for the calculations
is a self-consistent Keldysh Green's function approach
as described in Ref.~\onlinecite{kemper_effect_2014}

We couple the electrons in this band to a spectrum of bosons characterized by an Eliashberg function $\alpha^2 F(E)$ for a spectrum of modes with frequency $\Omega_{\gamma}$, which is
\begin{align}
\alpha^2 F(E) &= \sum_{\gamma} g_{\gamma}^2 \delta(E- \hbar \Omega_{\gamma}).
\end{align}
In particular, we use a dominant boson mode centered at $\hbar \Omega =$ 0.075 eV, with $g^2 = 0.0065 \si{eV\squared}$ corresponding to dimensionless coupling $\lambda = 0.2$, extracted from the slope of the real part of the retarded equilibrium self-energy at zero energy:
\begin{align}
\lambda \equiv -\frac{\partial \text{Re} \Sigma(E)}{\partial E} \Big|_{E=0}.
\end{align}
To ensure finite relaxation times at low energy, we add a second weakly coupled mode centered at 0.007 eV with $g^{2} = 2 \times 10^{-5} \si{eV\squared}$. The frequency of the strongly coupled bosonic mode was chosen to match the experimentally determined position of the crossing point of fluence- and energy-dependent decay times, as well as the nodal kink position. The coupling strength was adjusted to give roughly the correct time scale in the low-fluence limit for $E > \hbar \Omega$.

We excite this metallic single-band model system by a spatially homogeneous pump pulse, whose vector potential couples to the band electrons through the Peierls substitution $\kk\rightarrow \kk-e\textbf{A}(t)$ where $\kk$ is wavevector, $e$ is the electron charge and $\textbf{A}(t)$ is the time-dependent vector potential. (Here the speed of light $c=1$.) The vector potential ${\bf A}(t)$, which is in the Hamiltonian gauge, points along the zone diagonal and has a temporal shape with a Gaussian envelope ($\sigma =$ 20 fs), oscillation frequency $\hbar \omega =$ 0.35 eV, and peak strength
$A_{\text{max}}$. For the simulations in this work, we use field strengths
$A_{\text{max}} =$ 0.05, 0.15, 0.25, and 0.50 in units of V/$a_0$. For the given parameters and a typical lattice constant $a_0 = 3.8 \text{\AA}$ (Cu-Cu distance), these field strengths correspond to peak electric fields of 0.65, 2.0, 3.3, and 6.5 MV cm$^{-1}$, respectively.
The time-resolved ARPES signal is obtained in a postprocessing step using the formalism described in Refs.~\onlinecite{freericks_theoretical_2009, sentef_examining_2013} with probe pulse of width $\sigma_{\text{pr}} =$ 40 fs. The energy-resolved intensity is obtained by integrating the momentum- and energy-resolved data along the
Brillouin zone diagonal cut.
The decay times shown in the main text (Fig.~4d-f) are extracted using fits to the normalized intensity changes, binned in energy windows of 20 meV width like the experimental data and convolved with a Gaussian envelope function of full width XC to render the theoretical results directly comparable to the experiment.



The present modeling and analysis are based mostly on the energy scale ($70-75$ $\si{meV}$), extracted from the nodal kink position (below $E_{F}$) as well as the reported population decay time step position (above $E_F$). Importantly, statements about coupling to bosonic modes with energies higher than $100$ $\si{meV}$ cannot be drawn from our analysis.

\subsection{Theoretical comparison of e-e and electron-phonon scattering}

The e-e interactions are treated
self-consistently at second order with a local interaction strength $U_0$:
\begin{align}
\Sigma^\mathcal{C}(t,t') = -i^2 U_0^2 G^\mathcal{C}(t,t') G^\mathcal{C}(t,t') G^\mathcal{C}(t',t)
\end{align}
where $G^\mathcal{C}(t,t')$ is the double-time contour-ordered Green's function\cite{stefanucci_nonequilibrium_2013}.
Since the interaction strength depends on the filling, we define an effective $U:\ U \equiv U_0^2 n(1-n)$ for a better comparison
to the e-p coupling strength. For the comparison, we used a simplified tight-binding model with $t=0.25$ eV, $t'=0.075$
 eV and $\mu=-0.255$ eV.  We use a strongly coupled phonon with $\Omega=0.1$ eV and $g^2$=0.02 eV$^2$. It should be noted
 that the parameters for the comparison were chosen to make the calculations computationally feasible and make a clear distinction
between e-e and e-p scattering, rather than for a quantitative comparison to data (as in the other sections of the manuscript).

\section*{$\tau_{\mathrm{long}}$ versus $E_{\mathrm{bin}}$ and $\Phi$}

\begin{figure}
  \includegraphics[scale=.75]{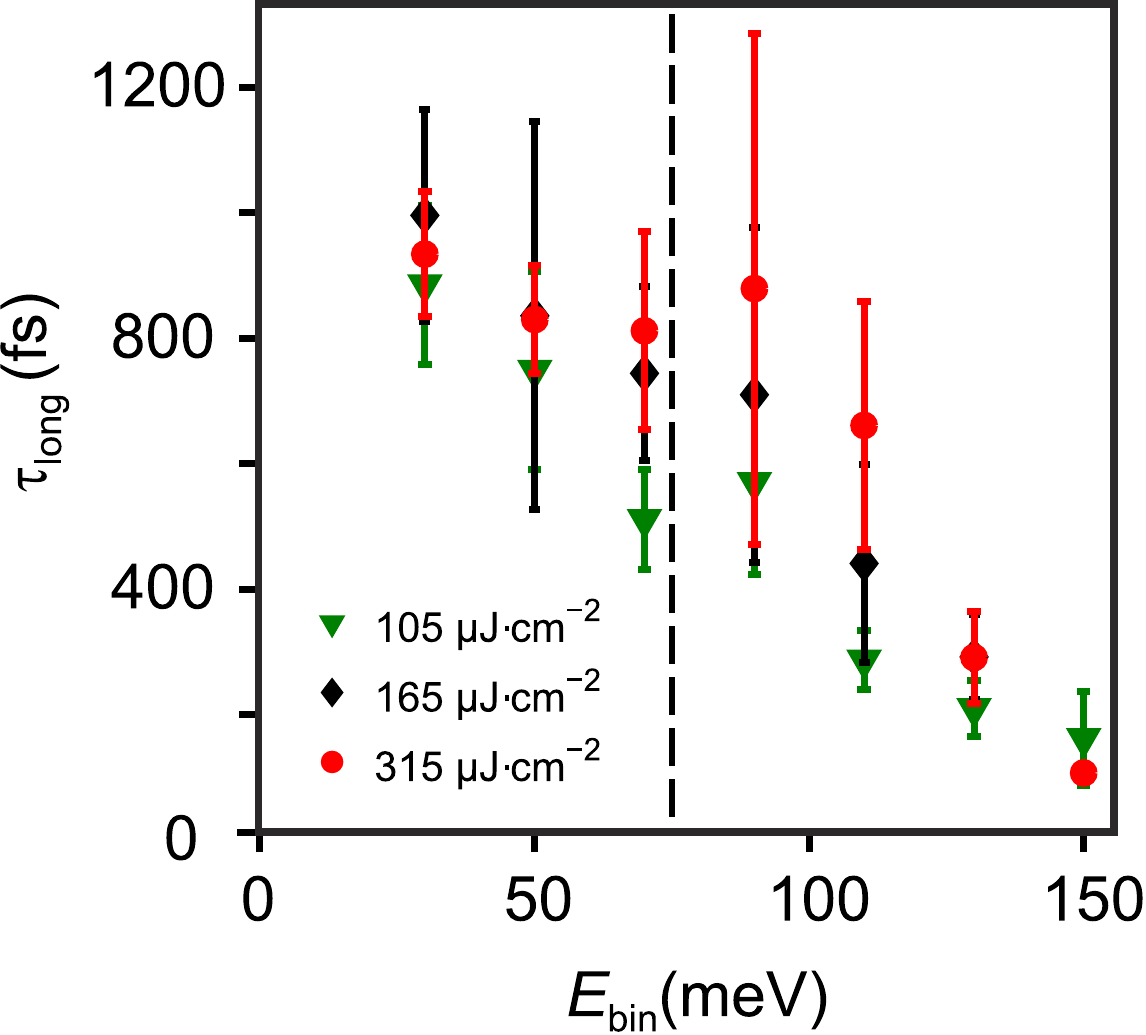}\\
  \caption{$\tau_{\mathrm{long}}$ versus $E_{\mathrm{bin}}$, as defined in the main text, for $\Phi =$ 105 (green) 165 (black) and 315 (red) $\mathrm{\mu J \cdot cm^{-2}}$. Note the absence of the effect of any energy scale on $\tau_{\mathrm{long}}$. The vertical dashed line at 75 meV denotes the theoretical position of $+\hbar\Omega$.}\label{taulong}
\end{figure}

The long timescale $\tau_{\mathrm{long}}$, shown in Fig.~\ref{taulong} was found to carry no fluence dependence, within error. Interestingly, it was found to nearly merge with, or at least become indistinguishable, from $\tau_{\mathrm{short}}$ at high energies, though higher statistics and time resolutions would be required to ascertain this with certainty. Regardless, $\tau_{\mathrm{short}}$ is found robustly at all fluences measured. The fluence independence of $\tau_{\mathrm{long}}$ plays an important role in our extraction of $\tau_{\mathrm{short}}$ at the lowest fluence, particularly around $E_{\mathrm{bin}} = 70$ meV. Because at $\Phi = 35 \mathrm{ \mu J\cdot cm^{-2}}$ the statistics are relatively low as $E_{\mathrm{bin}}$ increases much past 70 meV and at later times, consistent fitting requires $\tau_{\mathrm{long}}$ to be held constant from higher fluence data. Relaxing this constraint of course does not alter the trends for $\tau_{\mathrm{short}}$ visible in the raw data, nor does it appreciably alter the results for the lowest fluence in e.g. Fig. 4a of the main text viz a viz the step around 70 meV. It does introduce a bit more uncertainty into the error bars but the position of the step remains.

\section*{Acknowledgments}

This work was supported in part by National Science Foundation
Grant No. PHYS-1066293 and the hospitality of the Aspen Center for
Physics. A.F.K. was supported by the Laboratory Directed Research
and Development Program of Lawrence Berkeley National Laboratory
under U.S. Department of Energy Contract No. DE-AC02-05CH11231.
J.K.F. was supported by the Department of Energy, Office of Basic
Energy Sciences, Division of Materials Sciences and Engineering
(DMSE) under Contract No. DE-FG02-08ER46542, and by the McDevitt
bequest at Georgetown. Computational resources were provided by
the National Energy Research Scientific Computing Center supported
by the Department of Energy, Office of Science, under Contract No.
DE-AC02-05CH11231. Work at Brookhaven National Laboratory was
supported by the Center for Emergent Superconductivity, an Energy
Frontier Research Center, headquartered at Brookhaven National
Laboratory and funded by the U.S. Department of Energy, under
Contract No. DE-2009-BNL-PM015. We acknowledge further funding
from the Deutsche Forschungsgemeinschaft through SFB 616 and SPP
1458, from the Mercator Research Center Ruhr through Grant No.
PR-2011-0003, and from the European Union within the seventh
Framework Program under Grant No. 280555 (GO FAST).

\end{document}